\input{aipcheck}

\documentclass[
    ,final            
  ]
  {aipproc}

\layoutstyle{6x9}

\newcommand{\be}{\begin{equation}}
\newcommand{\ee}{\end{equation}}
\def\kt{k_\perp}
\newcommand{\pup}{p^\uparrow}

\begin{document}

\title{Sivers distribution functions and the latest SIDIS data}

\author{M.~Anselmino}{
  address={Dipartimento di Fisica Teorica, Universit\`a di Torino,
             Via P.~Giuria 1, I-10125 Torino, Italy}
,altaddress={INFN, Sezione di Torino, Via P.~Giuria 1, I-10125 Torino, Italy}}

\author{M.~Boglione}{
  address={Dipartimento di Fisica Teorica, Universit\`a di Torino,
             Via P.~Giuria 1, I-10125 Torino, Italy}
,altaddress={INFN, Sezione di Torino, Via P.~Giuria 1, I-10125 Torino, Italy}
}
\author{U.~D'Alesio}{address={Dipartimento di Fisica, Universit\`a di Cagliari, I-09042 Monserrato (CA), Italy}
,altaddress={INFN, Sezione di Cagliari, C.P.~170, I-09042 Monserrato (CA), Italy}}
\author{S.~Melis}{address={European Centre for Theoretical Studies in Nuclear Physics and Related Areas (ECT*)
              Villa Tambosi, Strada delle Tabarelle 286, I-38123 Villazzano, Trento, Italy}}

\author{F.~Murgia}{address={Dipartimento di Fisica, Universit\`a di Cagliari, I-09042 Monserrato (CA), Italy}
,altaddress={INFN, Sezione di Cagliari, C.P.~170, I-09042 Monserrato (CA), Italy}}

\author{A.~Prokudin}{
  address={Jefferson Laboratory, 12000 Jefferson Avenue, Newport News, VA 23606}
}

\begin{abstract}
 We present an extraction of the Sivers distribution functions
from the most recent experimental data of the HERMES and COMPASS 
experiments, assuming a negligible contribution of sea quark Sivers functions. 
\end{abstract}

\maketitle


%
%

In Ref.~\cite{Anselmino:2008sga}, we presented an extraction of the Sivers 
distribution functions based on a fit of SIDIS experimental data from the 
HERMES~\cite{Diefenthaler:2007rj} and COMPASS~\cite{:2008dn,Martin:2007au} collaborations.
Data from HERMES~\cite{Diefenthaler:2007rj} presented an unexpectedly large $A_{UT}^{\sin(\phi_h-\phi_S)}$ asymmetry for 
$K^+$ production, about twice as much as the analogous asymmetry for $\pi^+$.
Such a large asymmetry suggested an important role of the sea Sivers functions.
Our analysis confirmed this expectation finding a large contribution of the $\bar{s}$-Sivers function.
%

Since then, new experimental results have become available: the COMPASS collaboration has 
released new SIDIS data
off a proton 
target, showing a clear Sivers asymmetry~\cite{Alekseev:2010rw}; a new 
HERMES data analysis, based on a much larger statistics, while confirming 
the previous pion data, softens the enhanced peak in the $K^+$ Sivers 
azimuthal moment~\cite{:2009ti}. These new data prompted us to perform a new analysis. 
Here we present a preliminary phenomenological fit
of COMPASS~\cite{:2008dn,Martin:2007au} and HERMES~\cite{:2009ti} data
including only $u$ and $d$ valence quarks
in order to see if the new data can be described consistently
without any contribution of the sea Sivers functions.

\begin{figure}
  \includegraphics[height=.3\textheight,angle=-90]{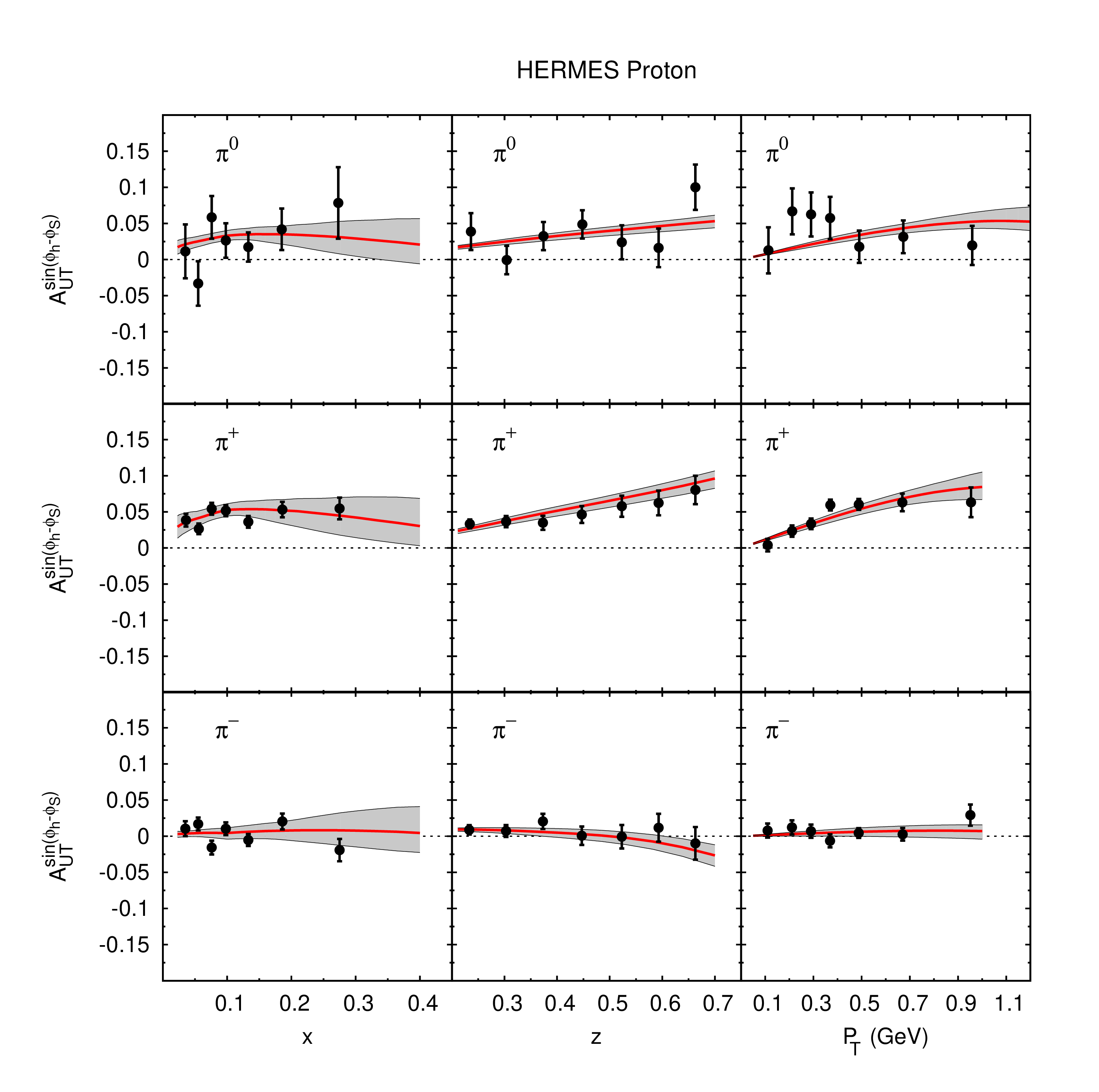}
 \includegraphics[height=.3\textheight,angle=-90]{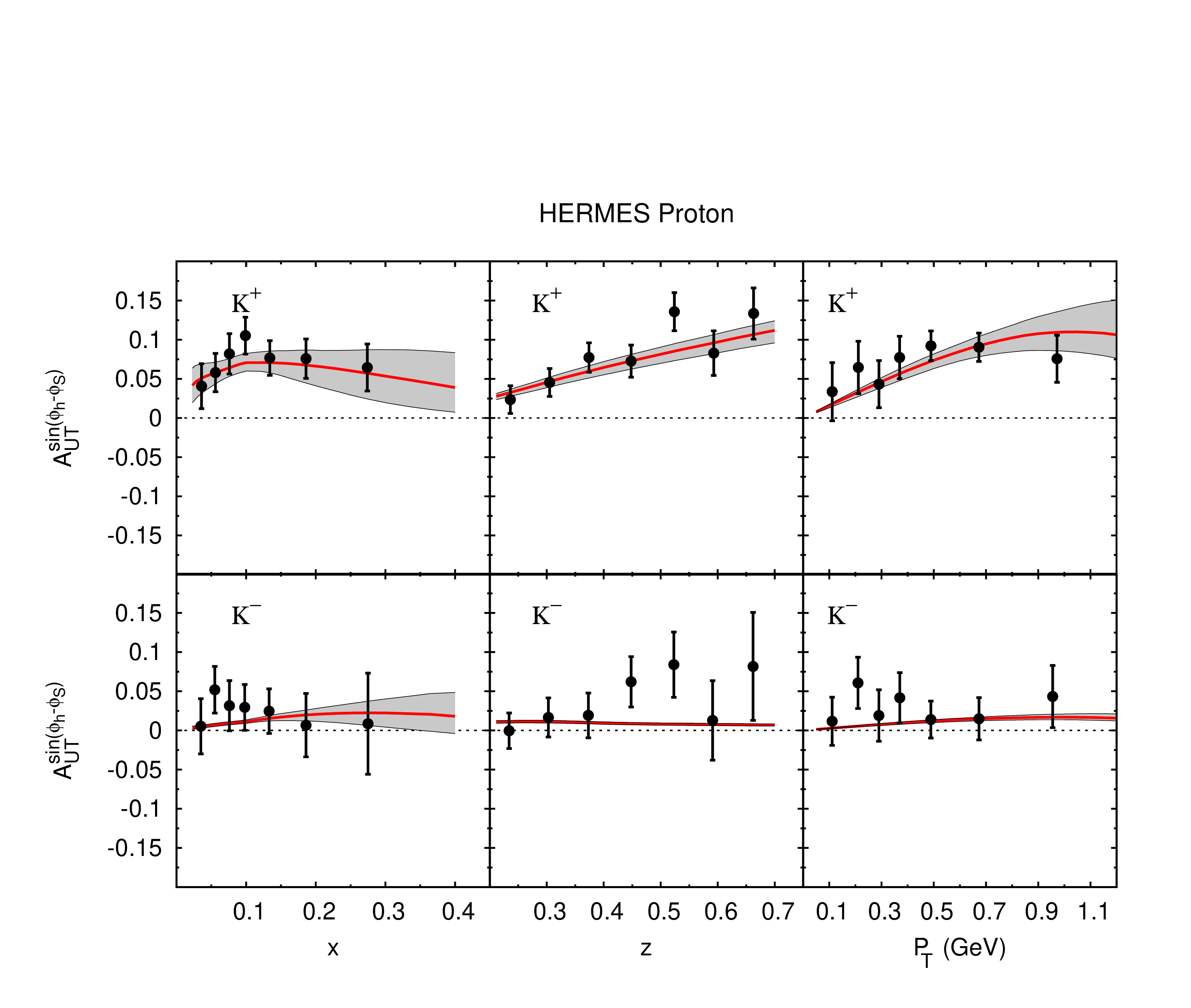}
\label{hermes-siv}
  \caption{Fit of HERMES data~\cite{:2009ti} for pion (left panel) and kaon production (right panel).}
\end{figure}

\begin{figure}
  \includegraphics[height=.3\textheight,angle=-90]{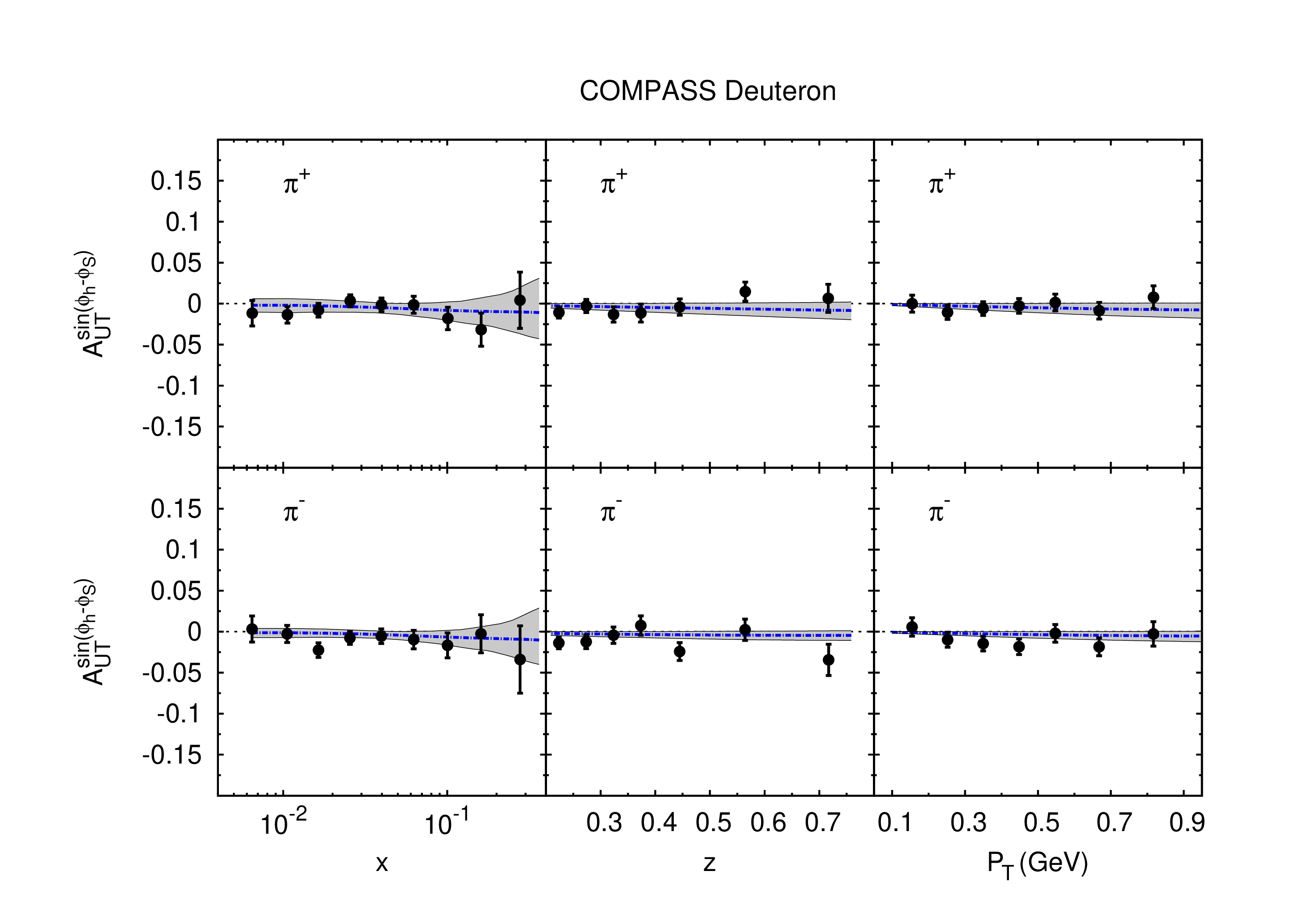}
 \includegraphics[height=.3\textheight,angle=-90]{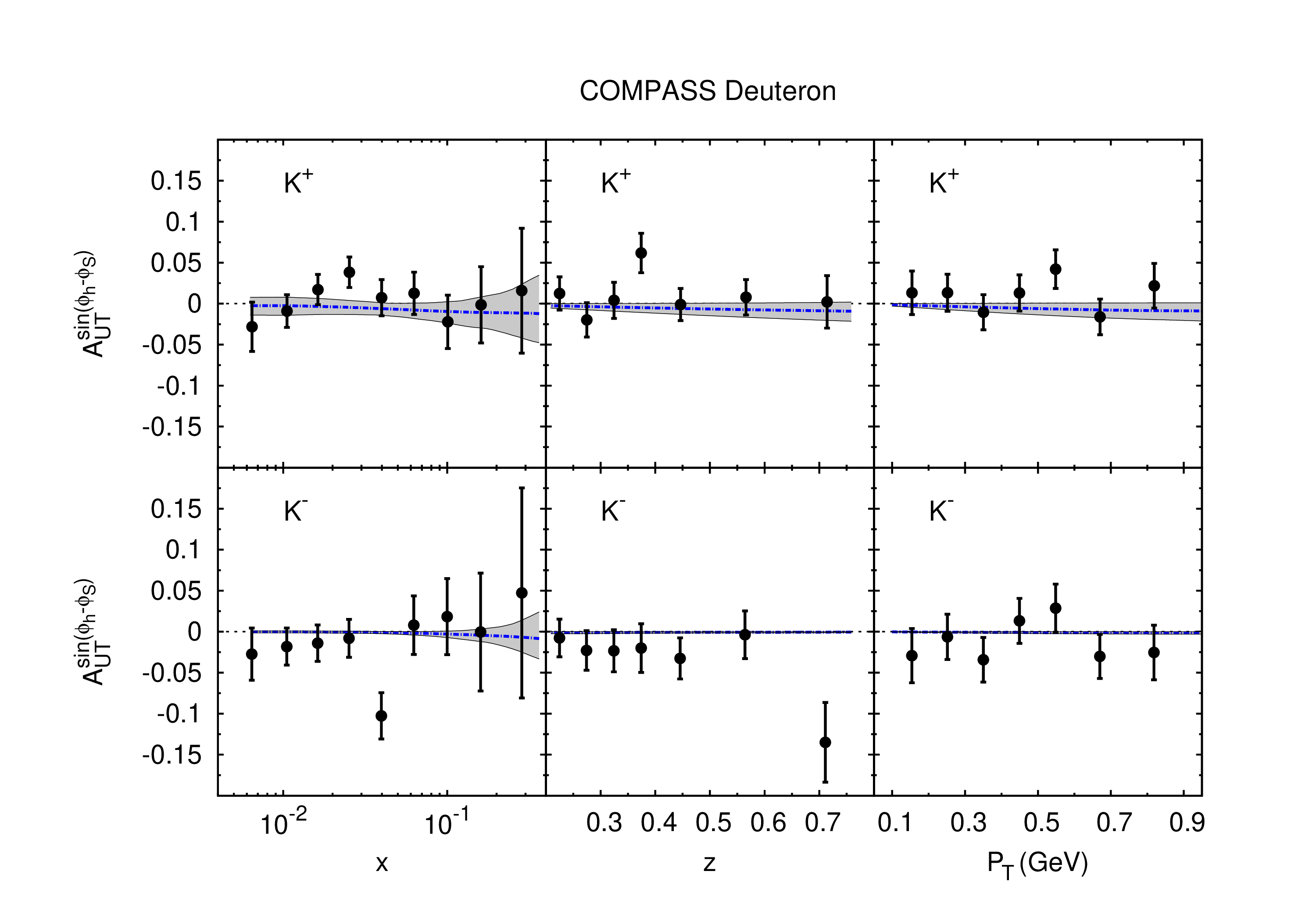}
\label{compd-siv}
  \caption{Fit of COMPASS deuteron data~\cite{:2008dn} for pion (left panel) and kaon production (right panel).}
\end{figure}

\begin{figure}
  \includegraphics[height=.3\textheight,angle=-90]{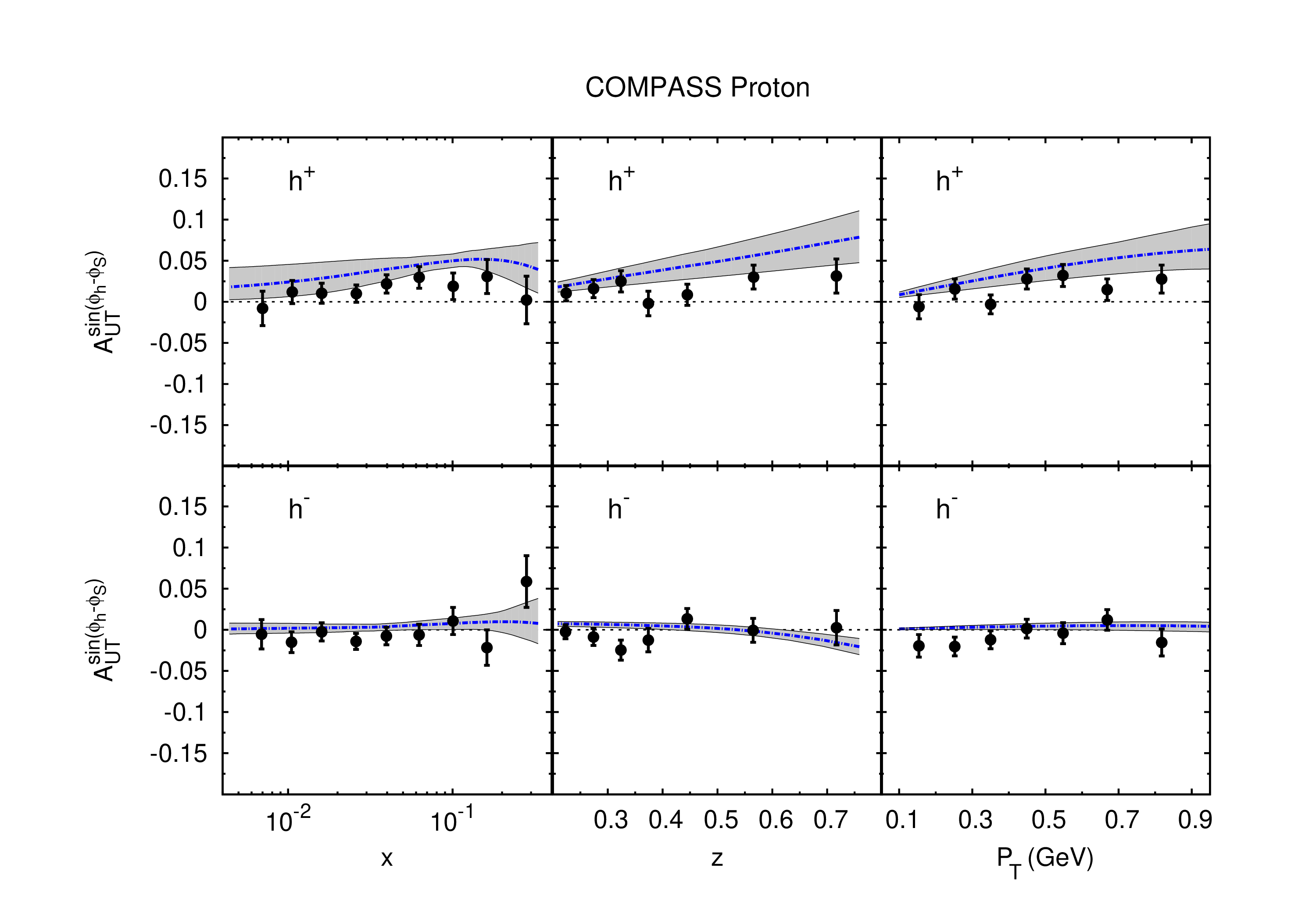}
 \includegraphics[height=.3\textheight,angle=-90]{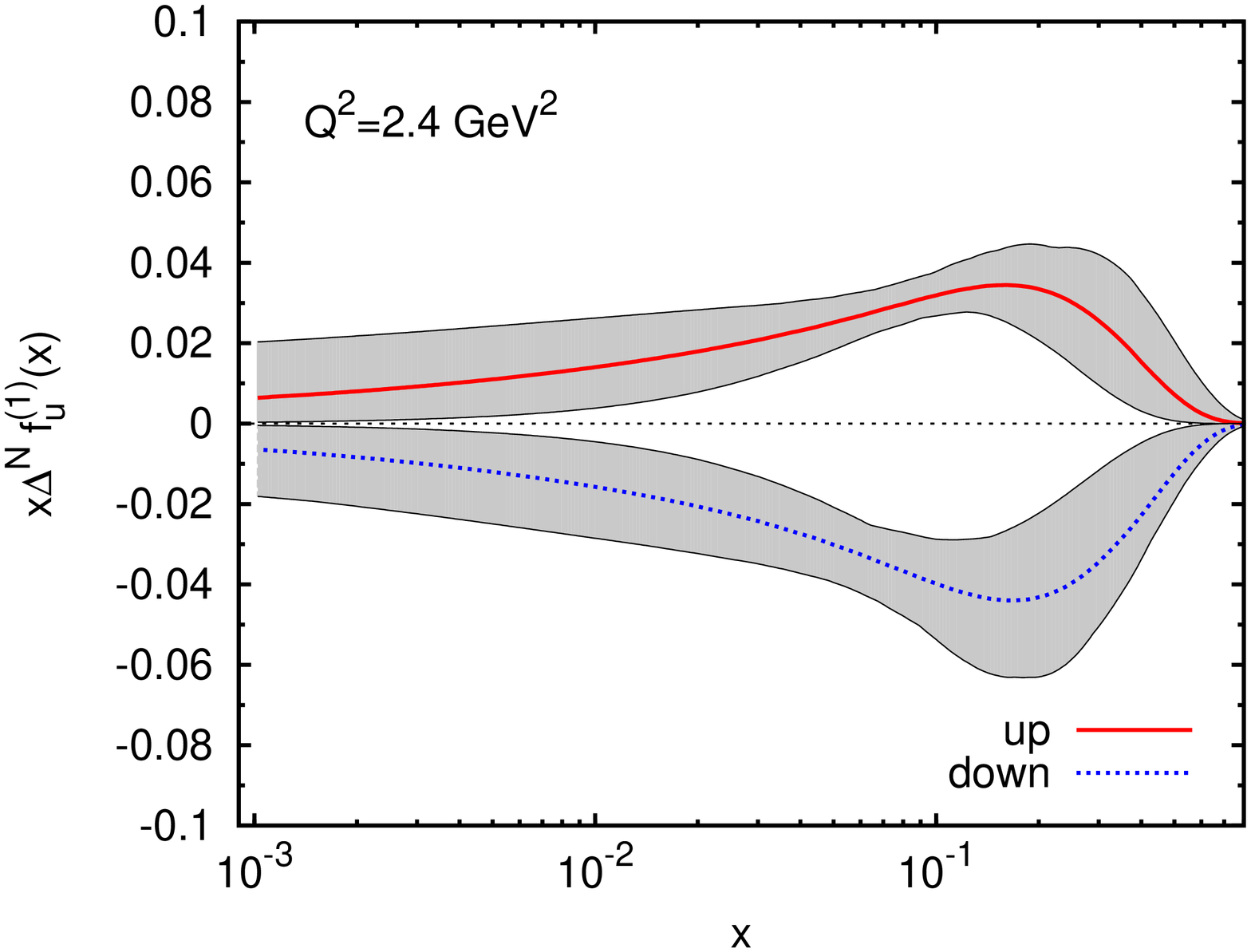}
  \caption{In the left panel we compare our predictions on proton target
for charged hadrons with the data released by the COMPASS collaboration~\cite{Alekseev:2010rw}.
The errors on the data are the statical and systematic errors added in quadrature.
The right panel shows the first moment of the Sivers functions extracted from the fitting procedure.}
\label{compp-siv-Sivers-fn}
\end{figure}

\begin{figure}
 \includegraphics[height=.25\textheight,angle=-90]{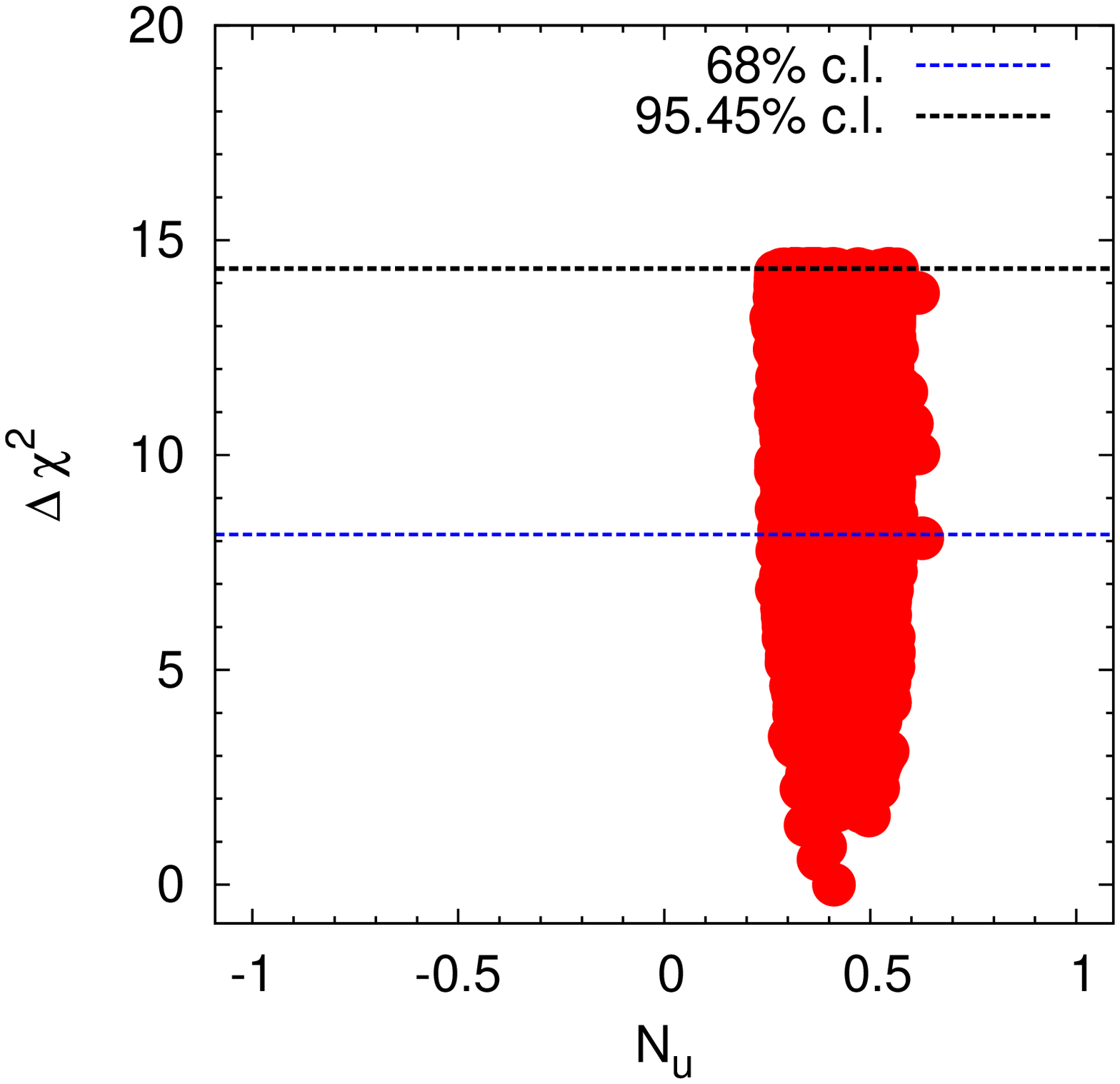}\hspace{-3pc}
\includegraphics[height=.25\textheight,angle=-90]{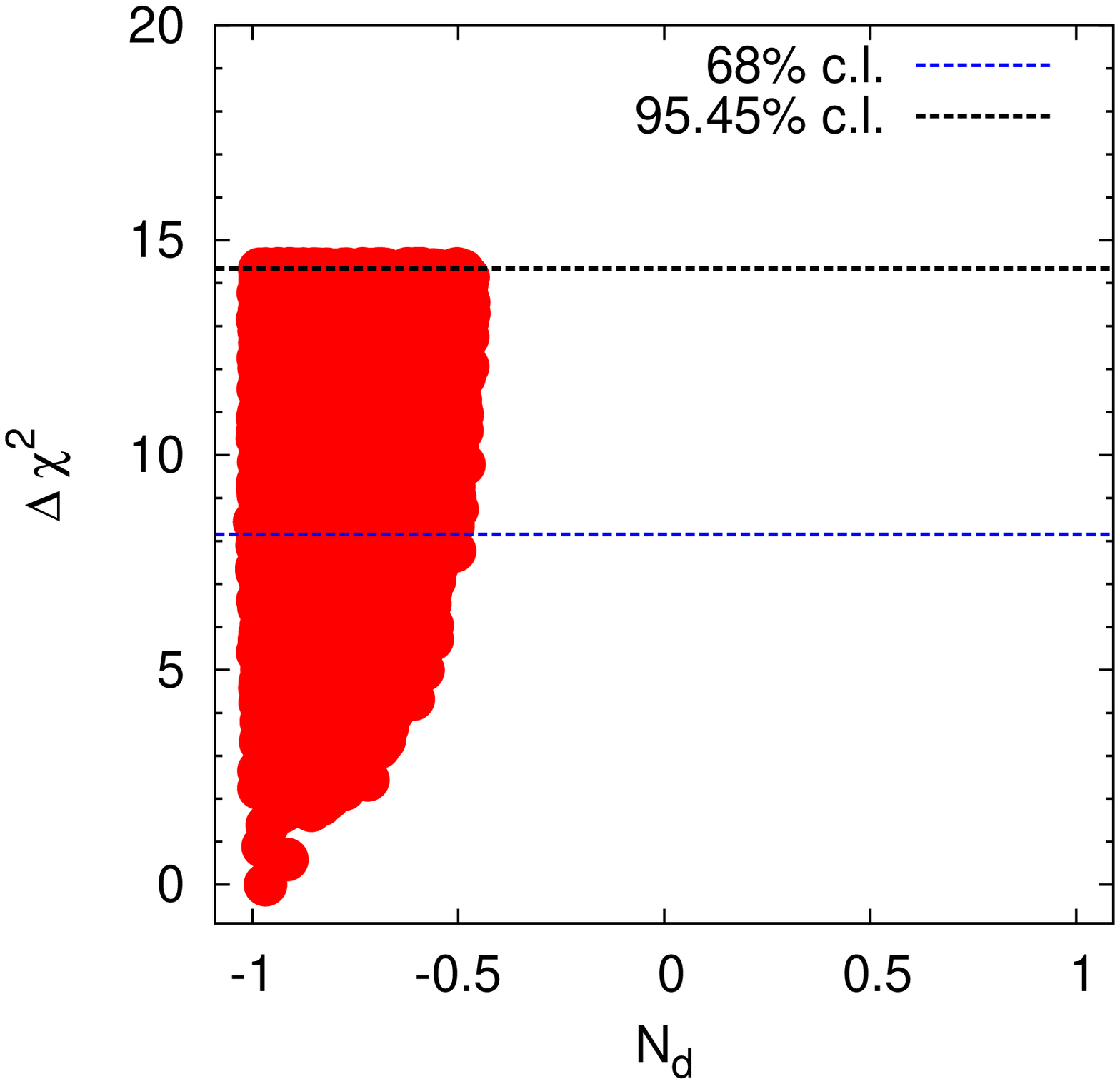} \hspace{-3pc}
 \includegraphics[height=.25\textheight,angle=-90]{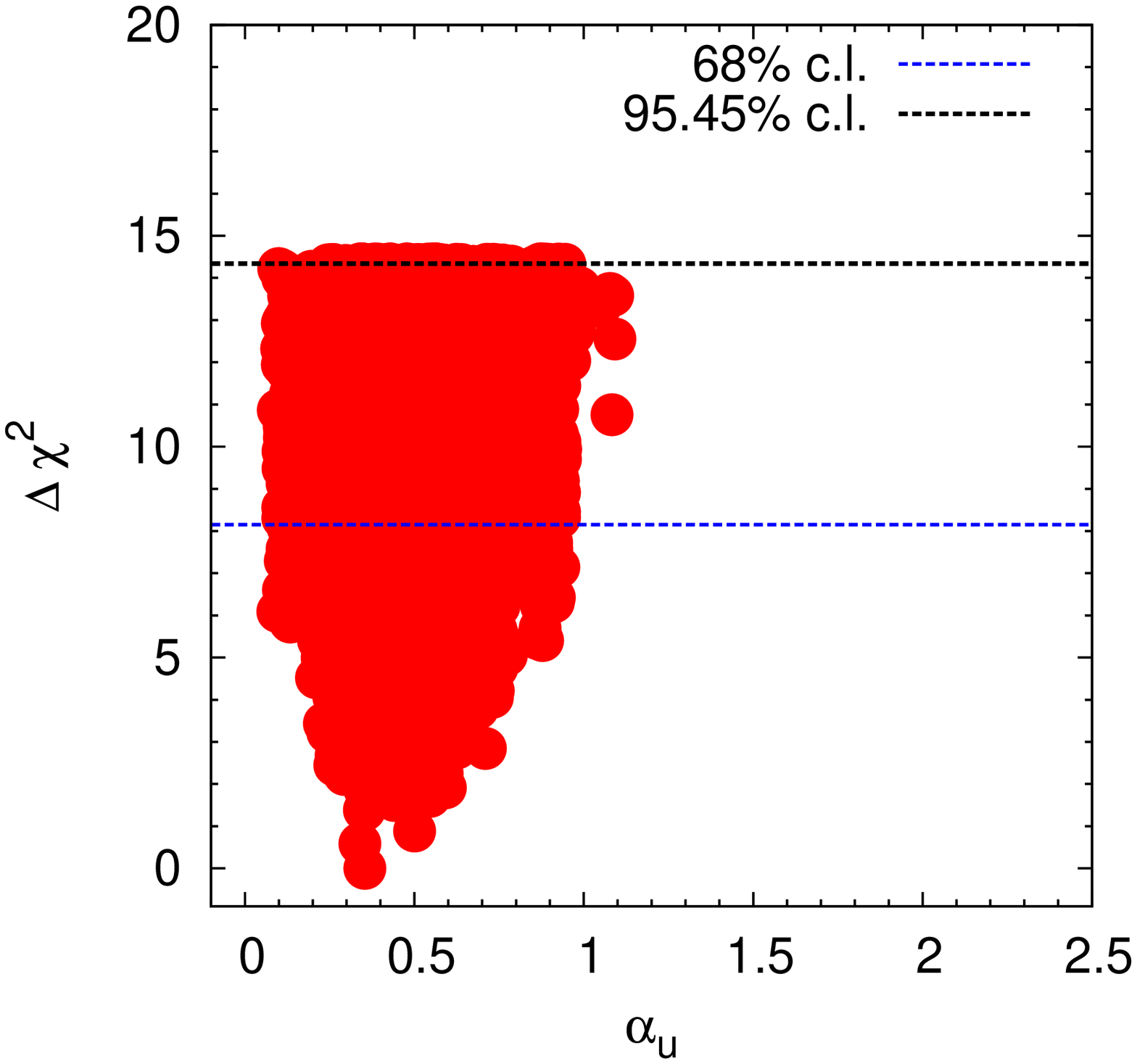}\hspace{-3pc}
\includegraphics[height=.25\textheight,angle=-90]{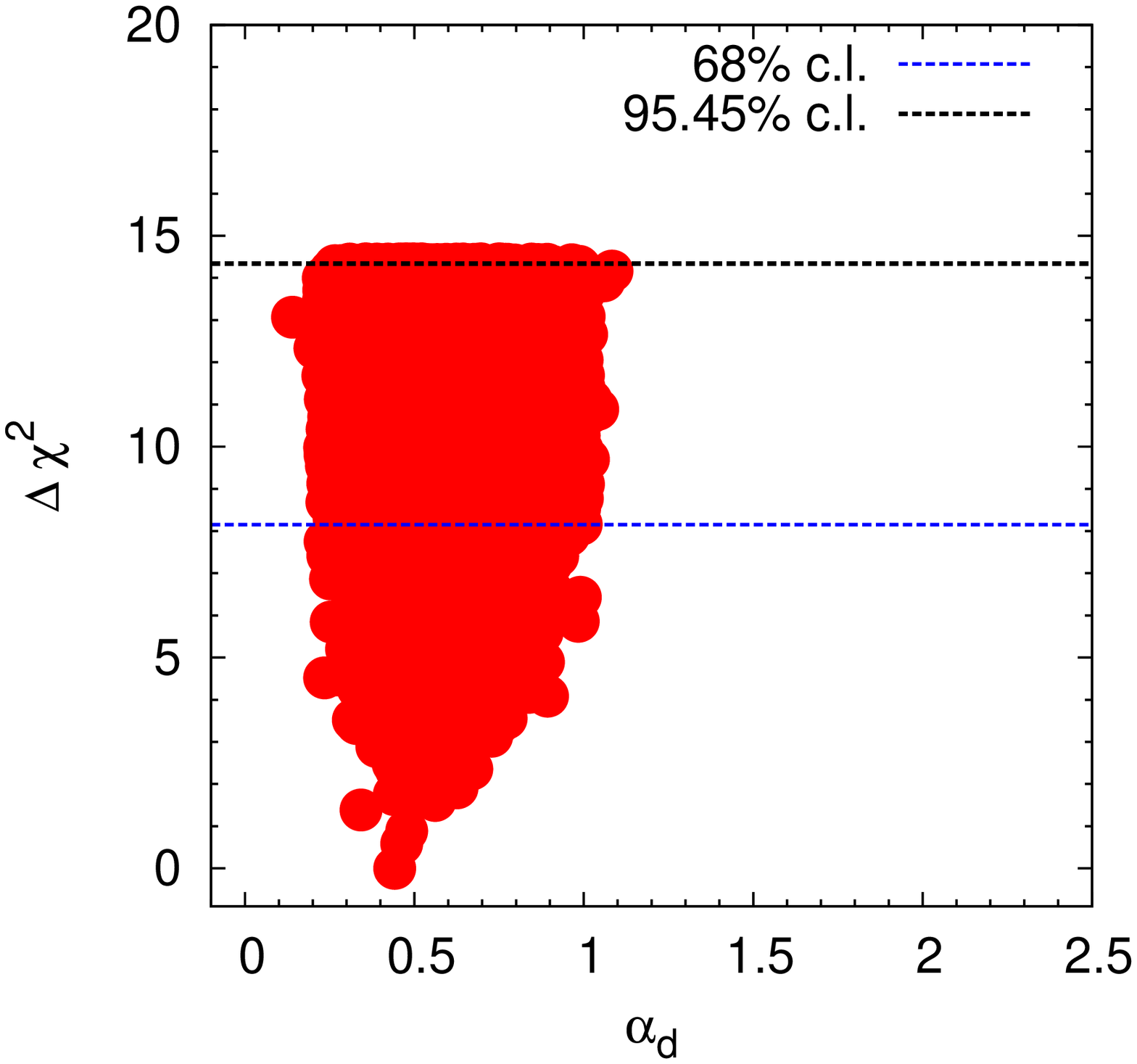} 
  \caption{From left to right: $\Delta \chi^2$ as a function of $N_u$, $N_d$, $\alpha_u$ and $\alpha_d$.}
\label{Sivers-par2}
\end{figure}

\begin{figure}
 \includegraphics[height=.25\textheight,angle=-90]{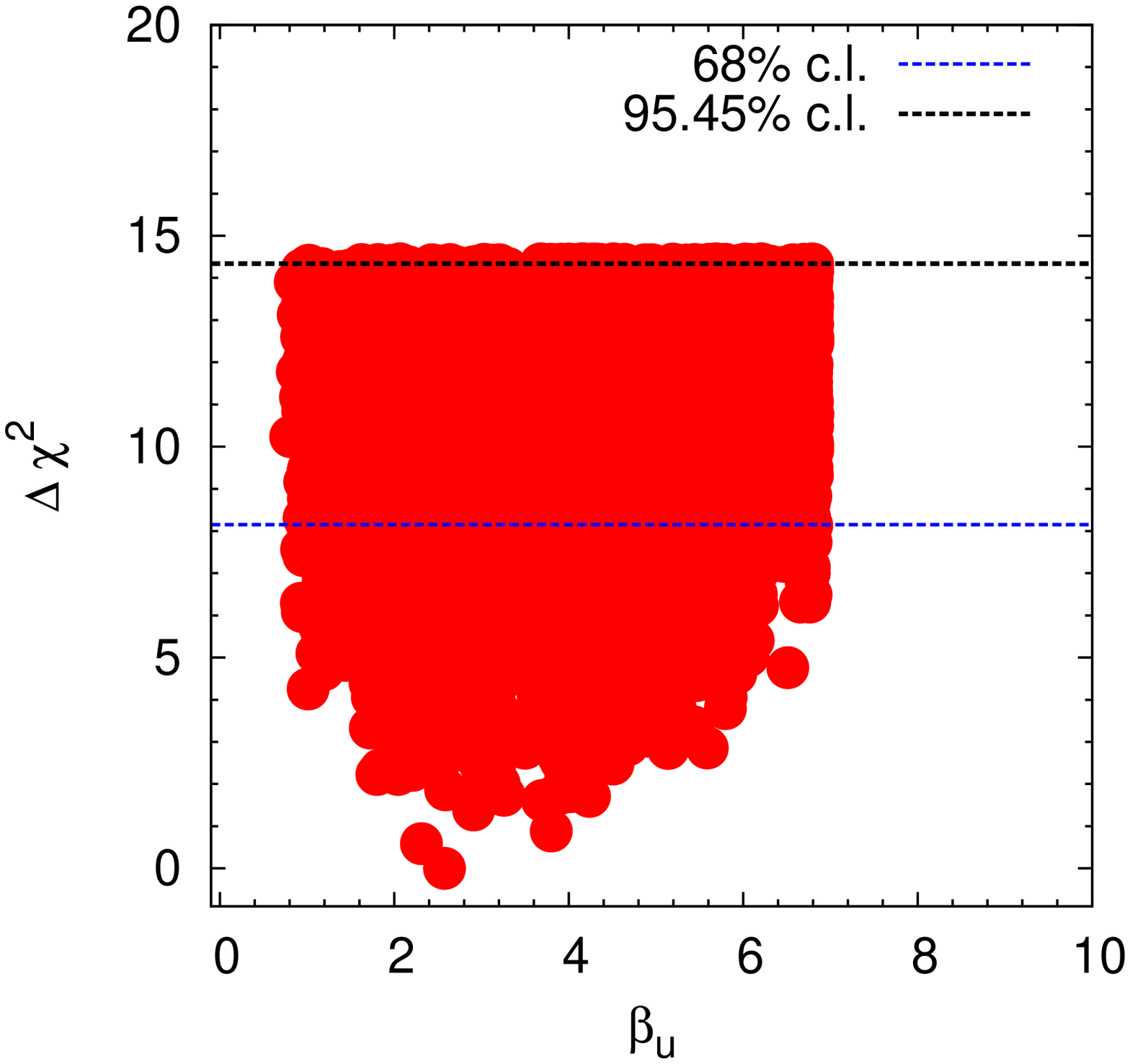}\hspace{-3pc}
\includegraphics[height=.25\textheight,angle=-90]{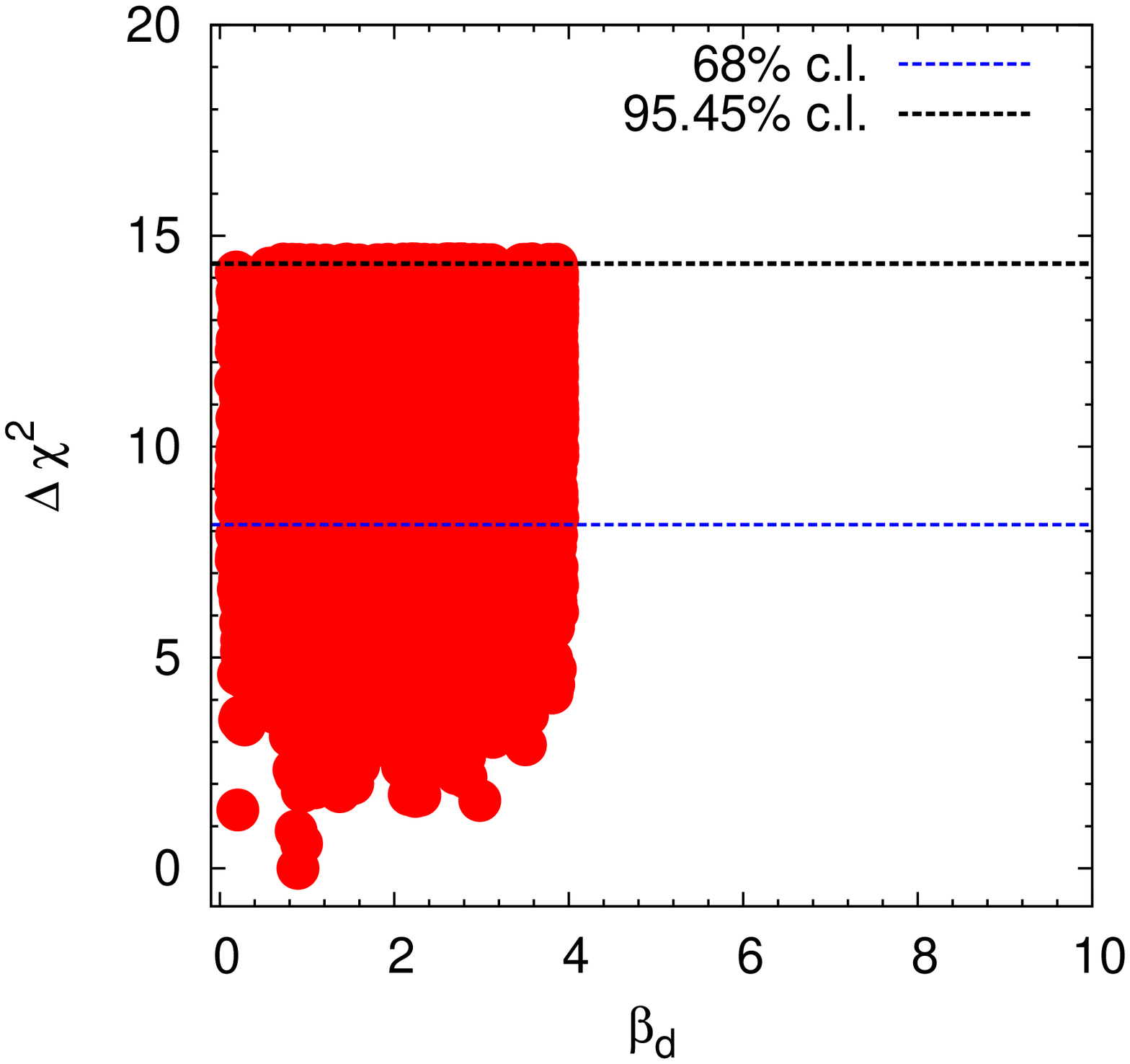}\hspace{-3pc} 

 \includegraphics[height=.25\textheight,angle=-90]{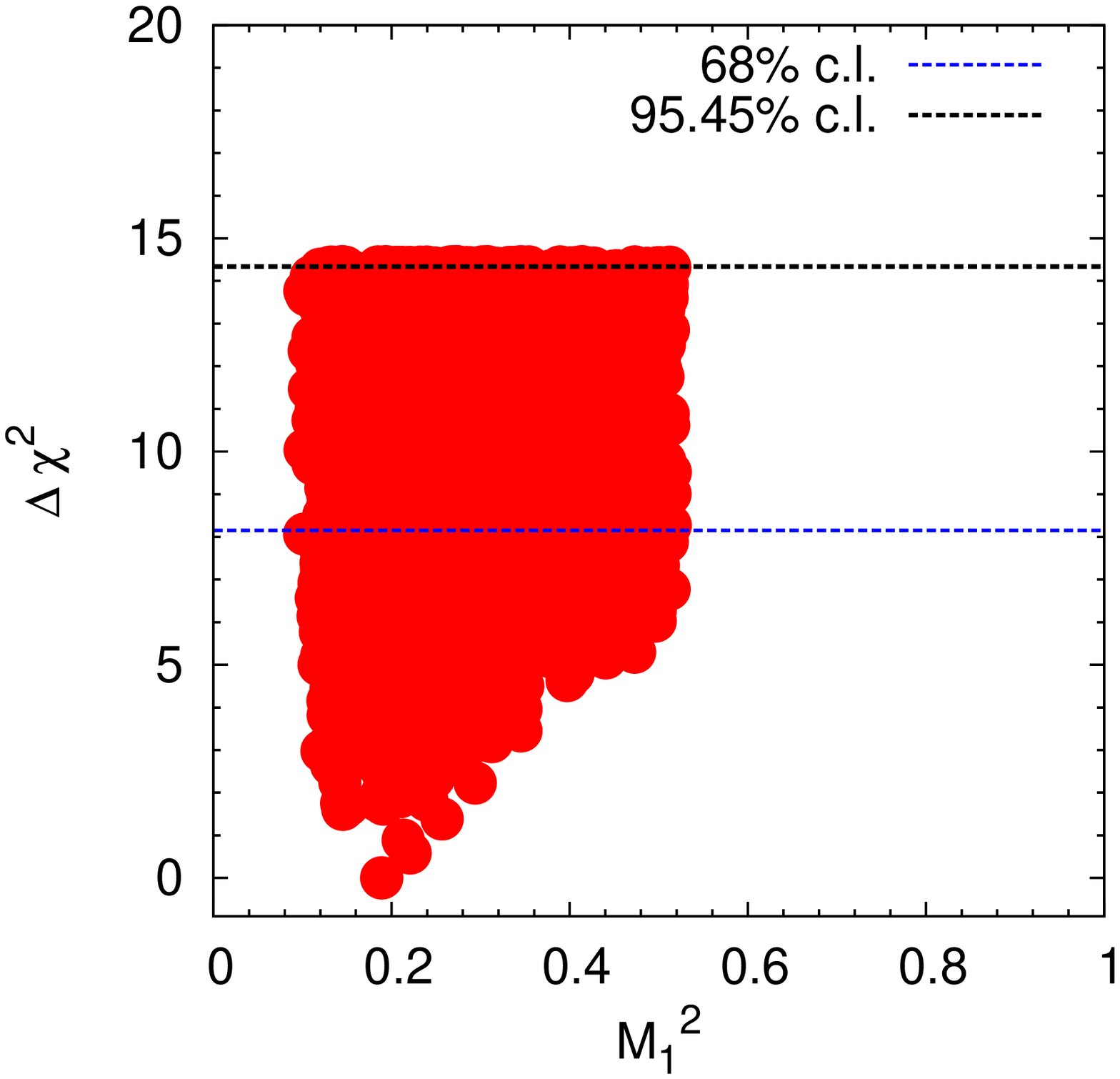}\hspace{-3pc}
\includegraphics[height=.25\textheight,angle=-90]{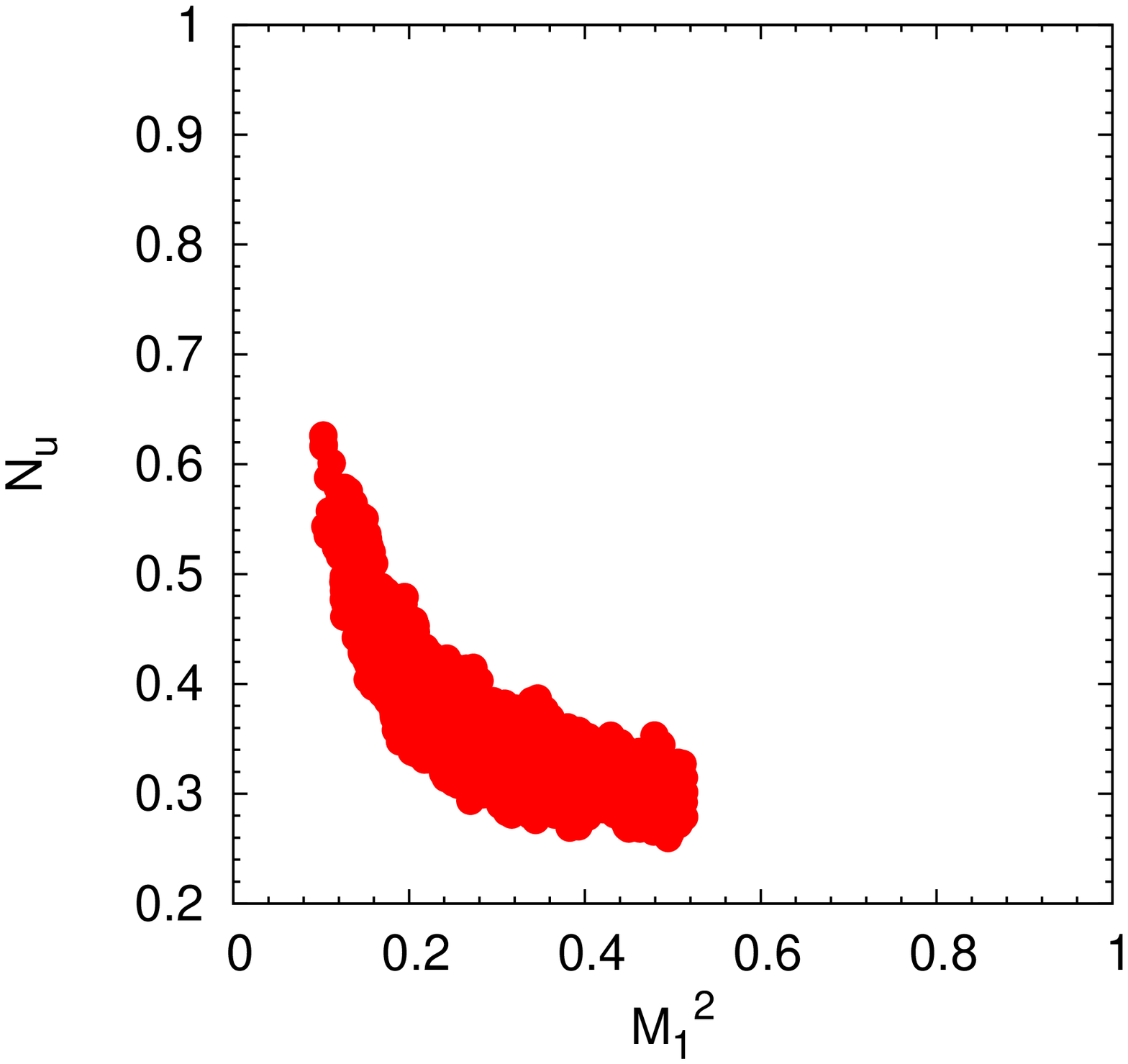} 
  \caption{From left to right: $\Delta \chi^2$ as a function of $\beta_u$, $\beta_d$, $M_1^2$
and correlation between $M_1^2$ and $N_u$.}
\label{Sivers-par4}
\end{figure}

The SIDIS transverse single spin asymmetry (SSA) 
$A^{\sin(\phi_h-\phi_S)}_{UT}$ measured by HERMES and COMPASS,
in the $\gamma^* - p$ c.m. frame and at order $k_\perp/Q$, is given 
by~\cite{Anselmino:2005ea, Anselmino:2005nn,Anselmino:2011ch}:
\be
A^{\sin (\phi_h-\phi_S)}_{UT} = 
\label{hermesut}
\frac{
\sum_q \!\int\!
{d\phi_S d\phi_h d^2 \textit{\textbf{k}}_\perp}
\Delta^N \! f_{q/\pup} (x,\kt) \sin (\varphi -\phi_S)
\frac{d \hat\sigma ^{\ell q\to \ell q}}{dQ^2}
 D_q^h(z,p_\perp) \sin (\phi_h -\phi_S) }
{
\sum_q \!\int \!{d\phi_S d\phi_h  d^2 \textit{\textbf{k}}_\perp}
f_{q/p}(x,k _\perp)  \frac{d \hat\sigma ^{\ell q\to \ell q}}{dQ^2} 
 D_q^h(z,p_\perp) } \> \cdot
\ee
where $\phi_S$ and $\phi_h$ are the azimuthal angles identifying the
directions of the proton spin $\textit{\textbf{S}}$ and of the momentum of the outgoing hadron $h$
respectively w.r.t. the lepton plane, while $\varphi$ defines the direction of the incoming
(and outgoing) quark transverse momentum,
$\textit{\textbf{k}}_\perp$ = $\kt(\cos\varphi, \sin\varphi,0)$;
$\frac{d \hat\sigma ^{\ell q\to \ell q}}{dQ^2}$ is the unpolarized
cross section for the elementary scattering  $\ell q\to \ell q$;
$D_q^h(z,p_\perp)$ is the fragmentation function describing the
hadronization of the final quark $q$ into the detected hadron $h$ with
momentum $\textit{\textbf{P}}_h$; $h$ carries, with respect
to the fragmenting quark, a light-cone momentum fraction $z$ and a transverse
momentum $\textit{\textbf{p}}_\perp$. Finally $\Delta^N \! f_ {q/\pup}(x,\kt)$
is the Sivers distribution function~\cite{Sivers:1990fh}:
\be
\Delta^N \! f_ {q/\pup}(x,\kt)=-\frac{2 k_{\perp}}{m_p}f_{1T}^{\perp}(x,k_{\perp})\,.
\ee
The Sivers function is parameterized in terms of the unpolarized
distribution function, as in Ref.~\cite{Anselmino:2005ea}, in the following
factorized form:
\be
\Delta^N \! f_ {q/\pup}(x,\kt) = 2 \, {\cal N}_q(x) \, h(\kt) \,
f_ {q/p} (x,\kt)\; , \label{sivfac}
\ee
with
\be
{\cal N}_q(x) =  N_q \, x^{\alpha_q}(1-x)^{\beta_q} \,
\frac{(\alpha_q+\beta_q)^{(\alpha_q+\beta_q)}}
{\alpha_q^{\alpha_q} \beta_q^{\beta_q}}\; ,
\hspace{2pc} {\rm and} \hspace{2pc} 
h(\kt) = \sqrt{2e}\,\frac{k_\perp}{M_{1}}\,e^{-{k_\perp^2}/{M_{1}^2}}\; ,
\ee
where $N_q$, $\alpha_q$, $\beta_q$ and $M_1$ (GeV/$c$) are free parameters
to be determined by fitting the experimental data. 
We adopt a Gaussian factorization for the unpolarized distribution and 
fragmentation functions with the Gaussian widths $\langle k_\perp^2\rangle$ 
and $\langle p_\perp^2\rangle$ fixed to the values found in 
Ref.~\cite{Anselmino:2005nn} by analysing the Cahn effect in unpolarized 
SIDIS: $\langle\kt^2\rangle = 0.25 \;({\rm GeV}/c)^2$ and 
$\langle p_\perp^2\rangle  = 0.20 \;({\rm GeV}/c)^2$.
For the unpolarized, $\kt$-integrated distribution and fragmentation functions
we use the GRV98~\cite{Gluck:1998xa} and DSS~\cite{deFlorian:2007aj} sets.
We best fit the HERMES proton and COMPASS deuteron data from 
Refs.~\cite{:2008dn,:2009ti} (209 points) including only Sivers functions for $u$ and $d$ quarks,
corresponding to seven free parameters, shown in table~\ref{tab:a}.
%
The results we obtain are rather satisfactory, with a 
$\chi^2_{dof}$ of about $1.06$. They are shown
in Figs.~\ref{hermes-siv}, \ref{compd-siv}. The corresponding Sivers functions 
are plotted in the right panel of Fig.~\ref{compp-siv-Sivers-fn}.
Such results are similar to those obtained in Ref.~\cite{Anselmino:2005ea}
The gray band in Figs.~\ref{hermes-siv}, \ref{compd-siv}, \ref{compp-siv-Sivers-fn} represents the statistical error
of the fitting procedure, calculated as in Ref.~\cite{Anselmino:2008sga}.
As shown in the left panels of Figs.~\ref{hermes-siv} and \ref{compd-siv},
pions data are well described by our fit.
The new HERMES data on kaon production
can be described reasonably well without any sea Sivers functions contribution,
although their inclusion in the fit procedure
can slightly improve the $\chi^2_{dof}$~\cite{Anselmino:2010bs}.
In the left panel of Fig.~\ref{compp-siv-Sivers-fn}
we show our predictions for the Sivers asymmetry at COMPASS kinematics on a proton target.
Although apparently we overestimate the positive charged hadron asymmetry,
COMPASS proton data are affected by a scale error of $\pm0.01$, not shown in the figure,
so that our predictions are still compatible with them. 
\begin{table}
\begin{tabular}{ccc}
\hline
&$\chi^2/dof=1.06$&\\
\hline
$N_u=0.40$ & $\alpha_u=0.35$ & $\beta_u=2.6$ \\
$N_d=-0.97$ & $\alpha_d=0.44$ & $\beta_d=0.90$\\
& $M_1^2=0.19\textrm{ GeV}^2$ &  \\
\hline
\end{tabular}
\caption{$\chi^2$ and best values of the parameters.}
\label{tab:a}
\end{table}
Figs.~\ref{Sivers-par2} and \ref{Sivers-par4} show the uncertainties of the fit parameters.
They are obtained following the procedure of Ref.~\cite{Anselmino:2008sga}.
We generate randomly new sets of parameters, then we collect the sets giving a variation of $\chi^2$
with respect to the minimum corresponding to a $95.45 \%$ confidence level.
The figures show the $\Delta \chi^2$ as a function of each parameter.
Some of the parameters are correlated.
The strongest correlation is between $N_u$ and $M_1^2$,
as it is shown in the most right panel of Fig.~\ref{Sivers-par4}.
It is interesting to 
notice that the $M_1$ parameter, which fixes the Gaussian width of the 
Sivers function (i.e. its distribution in $\kt$) is rather well constrained and turns out 
to be between one half and two thirds of the unpolarized distribution 
function width.

\bibliographystyle{aipproc}   

\bibliography{biblio}

\end{document}